\documentclass[a4paper,11pt]{article}
\usepackage{pos}
\usepackage{combelow}
\usepackage{subcaption}

\renewcommand{\d}{\mathrm{d}}
\newcommand{\epn}{\frac{\d E_\perp^{(0)}}{\d \eta}}

\title{System size dependence of pre-equilibrium and applicability of hydrodynamics in heavy-ion collisions}
\ShortTitle{Pre-equilibrium and applicability of hydro in HIC}
\author[a]{Victor E. Ambru\cb{s}}
\author[b]{S.~Schlichting}
\author*[b,c]{C.~Werthmann}

\affiliation[a]{Department of Physics, West University of Timi\cb{s}oara, Bd.~Vasile P\^arvan 4, Timi\cb{s}oara 300223, Romania}
\affiliation[b]{Fakultät für Physik, Universität Bielefeld, D-33615 Bielefeld, Germany}
\affiliation[c]{Incubator of Scientific Excellence-Centre for Simulations of Superdense Fluids, University of Wrocław, pl. Maxa Borna 9, 50-204 Wrocław, Poland}

\emailAdd{cwerthmann@physik.uni-bielefeld.de}

\abstract{We simulate the space-time dynamics of high-energy collisions based on a microscopic kinetic description, in order to determine the range of applicability of an effective description in relativistic viscous hydrodynamics. We find that hydrodynamics provides a quantitatively accurate description of collective flow when the average inverse Reynolds number $\mathrm{Re}^{-1}$ is sufficiently small and the early pre-equilibrium stage is properly accounted for. By determining the breakdown of hydrodynamics as a function of system size and energy, we find that it is quantitatively accurate in central lead-lead collisions at LHC energies, but should not be used in typical proton-lead or proton-proton collisions, where the development of collective flow can not accurately be described within hydrodynamics.}

\FullConference{HardProbes2023\\
 26-31 March 2023\\
 Aschaffenburg, Germany\\}

\begin{document}
\maketitle

\section{Introduction}

Hydrodynamics typically plays a central role in mordern simulation frameworks of hadronic collisions, but its applicability to early times or small systems is questionable, as by construction it requires a scale separation between microscopic dynamics and the total system size as well as some degree of equilibration. Nevertheless, the discovery of collective flow in small systems~\cite{Nagle:2018nvi} has led to many attempts of describing their time evolution in hydrodynamics. This calls for studies that assess the applicability to hydrodynamics by comparing its performance to other dynamical descriptions that feature weaker prerequisites in terms of system size and proximity to equilibrium.

Kinetic theory describes the system as a statistical distribution of microscopic constituents and therefore has an extended regime of applicability compared to hydrodynamics. We employ a simplified kinetic theory description of transverse flow observables in order to perform a quantitative assessment of the applicability of hydrodynamics as a function of system size, based on the development of transverse flow.~\cite{Ambrus:2022koq,Ambrus:2022qya}.

\section{Setup}
\subsection{Kinetic Theory}
We describe the system by a phase space distribution $f$ of massless on-shell bosons, assuming boost invariance and vanishing longitudinal momentum and transverse anisotropy at initial time. The time evolution is described via the Boltzmann equation in conformal relaxation time approximantion (RTA).
\begin{align}
        p^\mu \partial_\mu f =C_{\rm RTA}[f]=- \frac{p^\mu u_\mu}{\tau_R} (f-f_{\rm eq}) \ , \ \ \  \tau_R=5\frac{\eta}{s}T^{-1}
\end{align}

One advantage of this simple setup is that results will depend only on the initial state geometry and a single dimensionless parameter, the opacity $\hat{\gamma}$~\cite{Kurkela:2019kip}. It is a measure of the total interaction rate in the system and depends only on viscosity $\frac{\eta}{s}$, transverse size $R$ and energy scale $\epn$. 
\begin{align}\hat{\gamma} =\left(5{\frac{\eta}{s}}\right)^{-1}\left(\frac{1}{a\pi} {R}{\epn} \right)^{1/4}
\end{align}
Our initial condition was obtained as an average over events from the $30-40\%$ centrality class of Pb+Pb collisions at $5.02\;\mathrm{TeV}$  (see \cite{Borghini:2022iym} for details). This fixes $R$ and $\epn$, so we vary the opacity via $\frac{\eta}{s}$. However, due to the scaling of conformal RTA, this is equivalent to varying system size.

\subsection{Scaled Hydrodynamics}

It is important to set up the hydrodynamic simulations in such a way that they can be reasonably compared to kinetic theory, i.e. such that one can expect agreement when both descriptions are in equilibrium. A natural consequence is that the transport coefficients in hydrodynamics have to be matched to conformal RTA. 

The different behaviour of hydrodynamics in pre-equilibrium necessitates a rescaling of the hydrodynamic initial condition. This rescaling has been estabished in the case of absence of transverse expansion, where the system can locally be described by Bjorken flow. We implemented these modifications into the hydrodynamic code vHLLE~\cite{Karpenko:2013wva} and used this for our simulations.

\subsection{Hybrid Simulations}

Another way to circumvent the different behaviour of hydrodynamics in pre-equilibrium is to employ a hybrid framework, where the system is first evolved in kinetic theory and subsequently the energy-momentum tensor is used as input for a further hydrodynamic evolution. We chose the time of switching descriptions based on equilibration as determined by the inverse Reynolds number
\begin{align}
    \mathrm{Re}^{-1}=\left(\frac{6\pi_{\mu\nu}\pi^{\mu\nu}}{e^2}\right)^{1/2}\;,\label{eq:Reinvdef}
\end{align}
which locally measures the relative size of non-equilibrium contributions. We define a global analogue by taking the energy density weighted average over transverse space, denoted by $\langle \cdot \rangle_\epsilon$.
We thus require that $\langle\mathrm{Re}^{-1}\rangle_\epsilon$ must have dropped below a certain value before switching to hydrodynamics.

\section{Equilibration and development of flow}

\begin{figure}
    \centering
    \includegraphics[width=.99\linewidth]{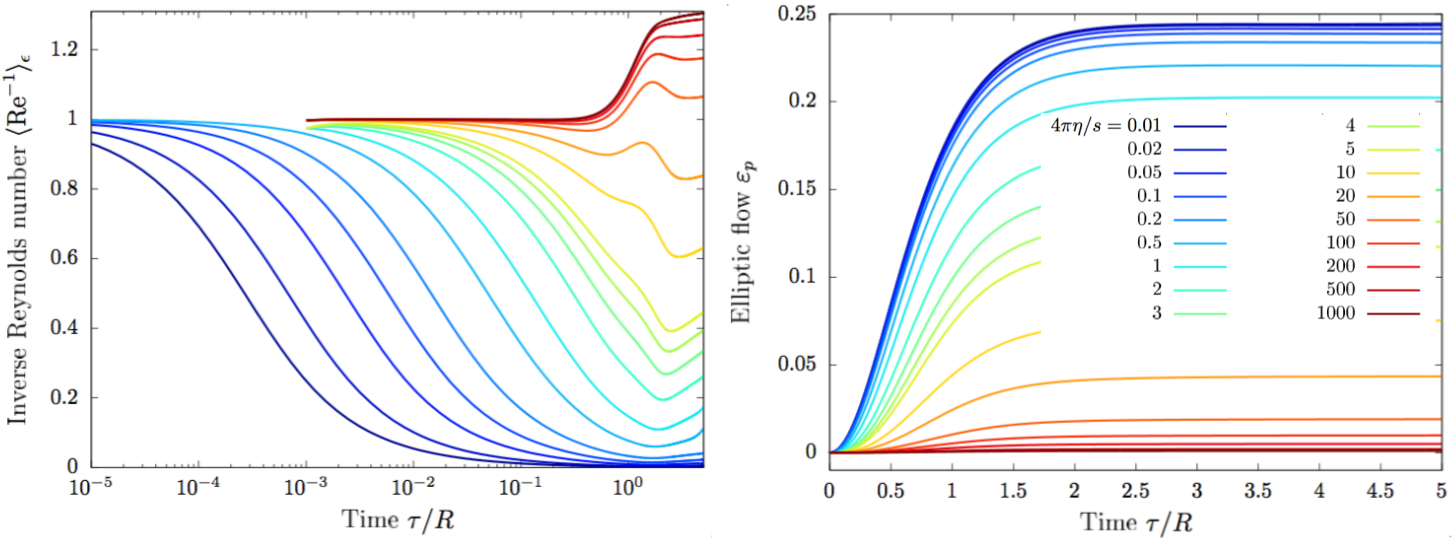}
    \caption{Time evolution of (left) inverse Reynolds number $\langle\mathrm{Re}^{-1}\rangle_\epsilon$ on a logarithmic scale and (right) elliptic flow $\varepsilon_p$. The color indicates the opacity of the respective system, as specified by the key.} 
    \label{fig:tevo_gdep}
\end{figure}

We first study the properties of different size systems as described in kinetic theory. The inverse Reynolds number $\mathrm{Re}^{-1}$ (see Eq.~\eqref{eq:Reinvdef}) quantifies the relative size of non-equilibrium effects in the system. It can be used to quantify how far away the system is from equilibrium. The left hand side of Figure~\ref{fig:tevo_gdep} shows the time evolution of this quantity for different opacities ranging over several orders of magnitude. It shows that the timescale of equilibration as indicated by the drop of $\mathrm{Re}^{-1}$ depends strongly on opacity. Transverse expansion setting in at $\tau\sim R$ will drive the system away from equilibrium and increase the value of $\mathrm{Re}^{-1}$, such that small systems never fully equilibrate.

The right hand side of Figure~\ref{fig:tevo_gdep} shows elliptic flow $\varepsilon_p$ as a function of time for the same opacities. $\varepsilon_p$ builds up on similar timescales $\tau\sim R$ for all system sizes. It continuously varies in magnitude from $\varepsilon_p = 0$ in the free-streaming case to a large opacity limit of $\epsilon_p \simeq 0.25$.

\section{Equilibration in hydrodynamics}

\begin{figure}
    \centering
        \centering
        \begin{subfigure}{.44\linewidth}
    \includegraphics[width=\linewidth]{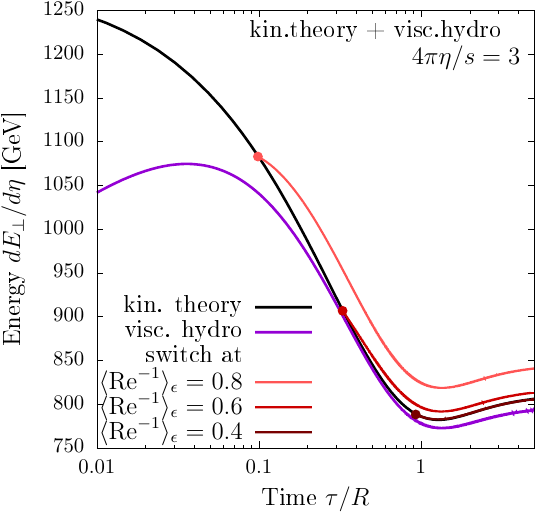}
        \end{subfigure}
        \hfill
        \begin{subfigure}{.54\linewidth}
            \includegraphics[width=\linewidth]{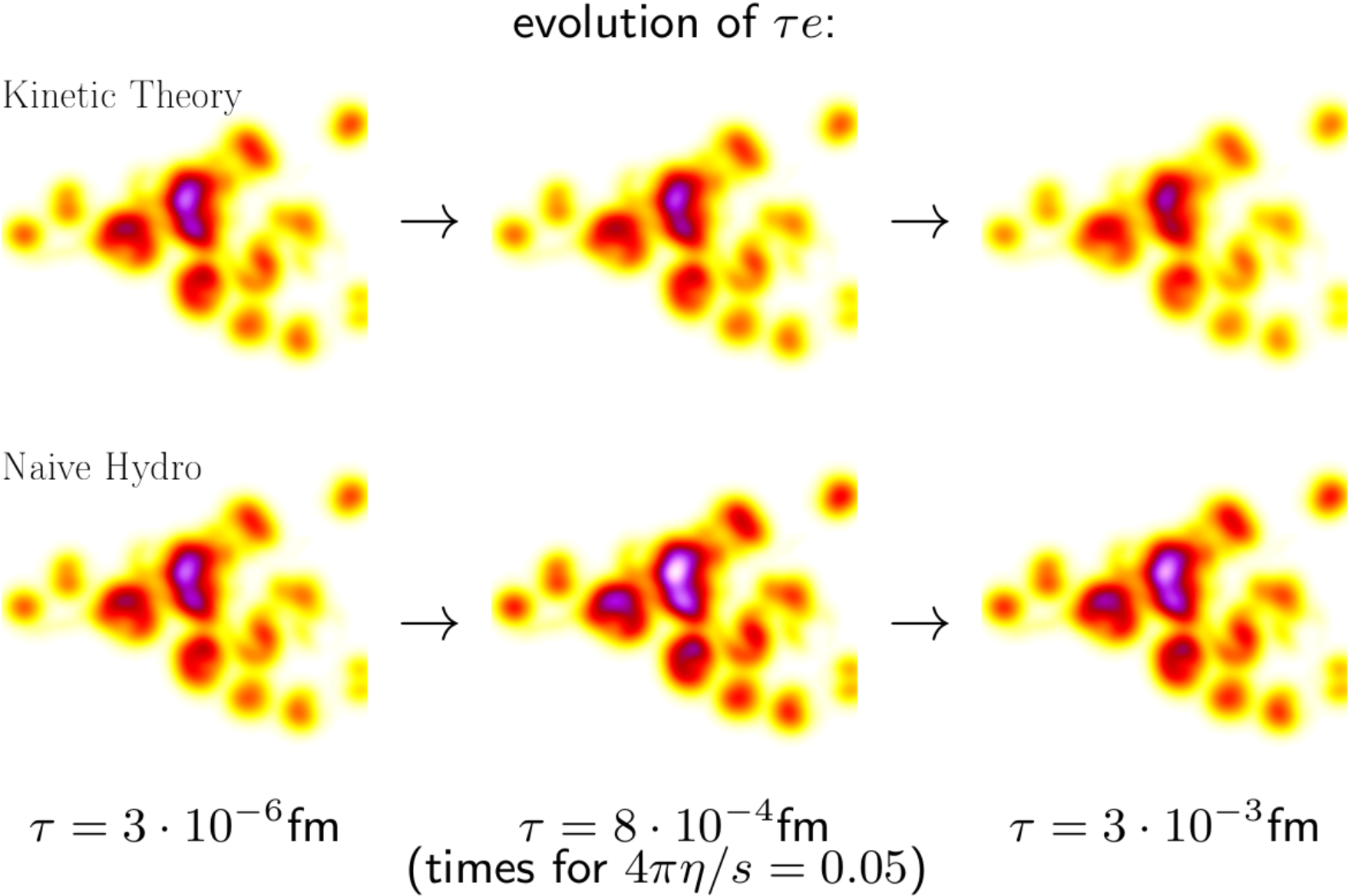}
            \vspace{8pt}
        \end{subfigure}
    \caption{Comparison of pre-equilibrium behaviour in kinetic theory and hydrodynamics via (right) the time evolution of an example profile of $\tau\epsilon$ and (left) the time evolution of transverse energy $\d E_\perp/\d\eta$. The right plot also shows time evolution in hybrid evolutions switching at different values of the inverse Reynolds number $\langle\mathrm{Re}^{-1}\rangle_\epsilon$.}
    \label{fig:hydro_preeq}
\end{figure}

As expected by construction, hydrodynamics shows deviations from kinetic theory during pre-equilibrium. If both models are initialized in the same way, this will cause them to be in disagreement when reaching equilibrium, as illustrated by the right hand side of Figure~\ref{fig:hydro_preeq}. The timescale of the equilibration dynamics depends on the local energy density. This means that cooling is inhomogeneous~\cite{Ambrus:2021fej}, which will cause a decrease of eccentricities by differing amounts in different dynamical descriptions.

We counteract this difference by applying a local scaling factor to the initial condition of hydrodynamics, which was calculated in Bjorken flow~\cite{Ambrus:2021fej}. Thus, the two descriptions have different initial conditions but come into agreement during equilibration. The left hand side of Figure~\ref{fig:hydro_preeq} shows the time evolution of transverse energy $\d E_\perp/\d \eta$ in different simulation setups. Hydrodynamics is initialized with a lower value, but dynamically comes into agreement with kinetic theory through the different pre-equilibrium behaviour. As the motivation for this initialization procedure relies on absence of transverse expansion, agreement will only be reached if the system reaches a sufficient degree of equilibration before the onset of transverse expansion. We also compare to hybrid simulations switching from a kinetic theory description of pre-equilibrium to hydrodynamics for late times, where switching times are defined based on $\mathrm{Re}^{-1}$. Early switching causes deviations due to the pre-equilibrium behaviour, meaning that if the system is closer to equilibrium when switching, the hybrid results are more accurate.

\section{Comparison of final state observables}

\begin{figure}
    \centering
    \includegraphics[width=0.49\linewidth]{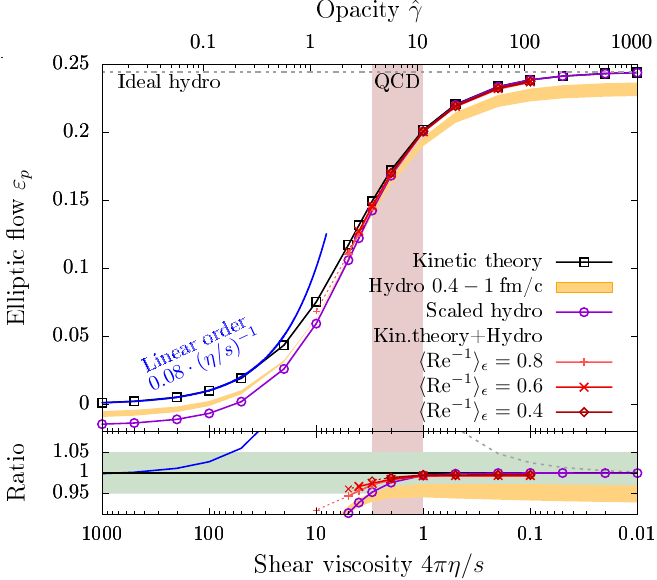}
    \includegraphics[width=0.49\linewidth]{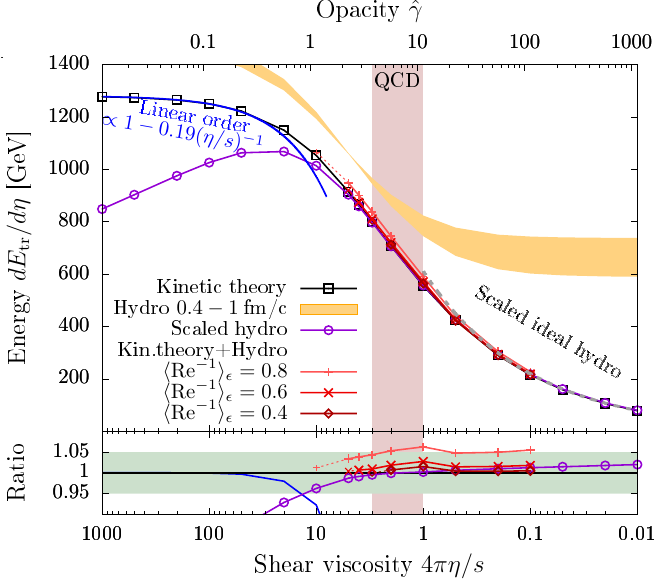}
    \caption{Final state values for (left) elliptic flow $\varepsilon_p$ and (right) transverse energy $\d E_\perp/d\eta$ as a function of opacity $\hat{\gamma}$ in different dynamical descriptuions. The red band indicates the physical values of $\eta/s$ in QCD. The lower part of the plots shows the ratio of each result to kinetic theory and the green band indicates the region of at most 5\% disagreement.}
    \label{fig:master}
\end{figure}

We want to compare the values of final state transverse flow observables as a basis for an assessment of the applicability of hydrodynamics. Figure~\ref{fig:master} shows results for the final state elliptic flow $\epsilon_p$ and transverse energy $\d E_\perp/\d \eta$. Naive hydrodynamics with unmodified initialization at $\tau_0=0.4 - 1\,$fm underestimates elliptic flow $\varepsilon_p$ and vastly overestimates transverse energy $\frac{\d E_{\rm tr}}{\d \eta}$. With our scaled initial condition, we can bring hydrodynamics into perfect agreement with kinetic theory at large opacities. The agreement remains reasonable down to $\hat{\gamma}\gtrsim 4$, which includes the physically relevant regime of QCD for this profile.
Hybrid results are also in good agreement at large opacities and can improve on scaled hydrodynamics at intermediate opacities. As seen before, when reducing the value of $\mathrm{Re}^{-1}$ at which hydrodynamics is turned on, one can expect better agreement with kinetic theory. If we require that the hybrid result should at most be in 5\% disagreement with kinetic theory, we can infer that hydrodynamics should only be applied after the inverse Reynolds number has dropped to a value $\langle \mathrm{Re}^{-1}\rangle_\epsilon \lesssim 0.75$.

\section{Regime of Applicability of hydrodynamics}

\begin{figure}
    \centering
    \includegraphics[width=.6\linewidth]{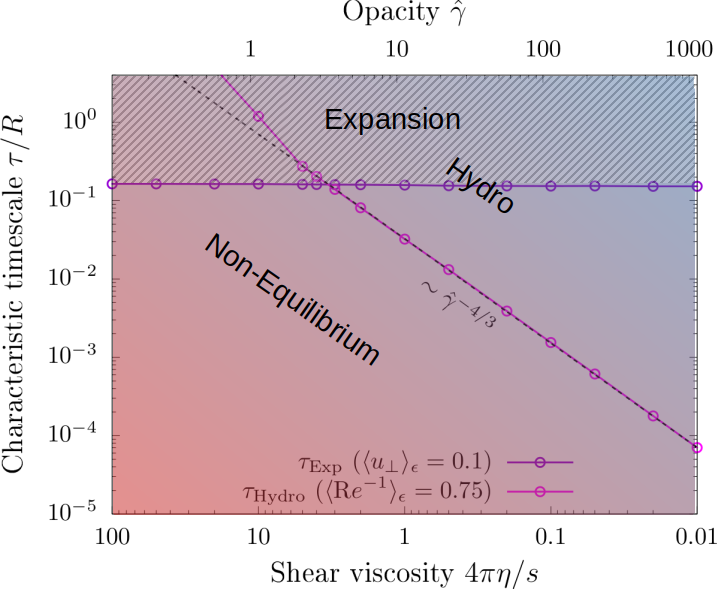}
    \caption{Diagram of the different dynamical regimes in the plane of evolution time $\tau$ and opacity $\hat{\gamma}$. The curves indicate the first times that (purple) transverse flow velocity reaches a value $\langle u_\perp \rangle_\epsilon$ indicative of the onset of transverse expansion and (pink) the inverse Reynolds number drops to $\langle\mathrm{Re}^{-1}\rangle_\epsilon=0.75$, meaning that hydrodynamics becomes applicable.}
    \label{fig:phase_diagram}
\end{figure}

 Figure~\ref{fig:phase_diagram} shows characteristic timescales in the evolution of the hadronic collision system as a function of opacity $\hat{\gamma}$. The two lines indicate the transition from non-equilibrium to hydrodynamic behaviour and from purely longitudinal to longitudinal plus transverse expansion.
 
 We define the timescale of the onset of transverse expansion by the first time transverse flow velocity reaches the value $\langle u_\perp\rangle_\epsilon =0.1$.  This timescale is mostly independent of opacity and takes values $\tau_{\rm Exp} \sim 0.2 R$. We also track our inferred criterion for the applicability of hydrodynamics, the drop in inverse Reynolds number to a value $\langle \mathrm{Re}^{-1} \rangle_\epsilon = 0.75$. This timescale follows a power law $\tau_{\rm Hydro}\propto \hat{\gamma}^{-4/3}$ before transverse expansion and depends strongly on system size.

We see that the curves representing these two criteria cross at an opacity value $\hat{\gamma}\sim 3$. For smaller opacities, hydrodynamics becomes applicable only after the onset of transverse expansion, if at all. Thus, a non-equilibrium description of transverse expansion is required. Estimates of the opacities show that hydrodynamics is not applicable in pp and pPb collisions, while central OO collisions just barely reach the regime of applicability~\cite{Ambrus:2022qya}.
\section{Conclusion}

We applied kinetic theory to the description of transverse flow on the full range in system size. When comparing to hybrid evolutions, we find that hydrodynamics becomes accurate on the 5\% level if $\mathrm{Re}^{-1}\lesssim 0.75$. This would mean that pp and pPb collisions are outside the regime of its applicability, but OO collisions cover the transition regime.

\end{document}